ARTICLE    OPEN

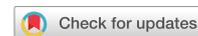

# Sign-reversible valley-dependent Berry phase effects in 2D valley-half-semiconductors

Xiaodong Zhou[1,2,6], Run-Wu Zhang[1,2,6], Zeying Zhang[3,6], Wanxiang Feng[1,2 ✉], Yuriy Mokrousov[4,5] and Yugui Yao[1,2]

Manipulating valley-dependent Berry phase effects provides remarkable opportunities for both fundamental research and practical applications. Here, by referring to effective model analysis, we propose a general scheme for realizing topological magneto-valley phase transitions. More importantly, by using valley-half-semiconducting $VSi_2N_4$ as an outstanding example, we investigate sign change of valley-dependent Berry phase effects which drive the change-in-sign valley anomalous transport characteristics via external means such as biaxial strain, electric field, and correlation effects. As a result, this gives rise to quantized versions of valley anomalous transport phenomena. Our findings not only uncover a general framework to control valley degree of freedom, but also motivate further research in the direction of multifunctional quantum devices in valleytronics and spintronics.

npj Computational Materials (2021)7:160 ; https://doi.org/10.1038/s41524-021-00632-3

## INTRODUCTION

Recent advances in valleytronics are mainly based on the paradigm of time-reversal-connected valleys, which generates valley polarization by an external field, dynamically or statically[1–3]. However, for achieving widespread applications of valleytronics, intrinsic properties are to be prioritized higher than external tunability. More importantly, intrinsic valleytronics materials hosting spontaneous valley polarization are most desirable, owing to their advantages of robustness, power efficiency, and simplicity in operation. In this regard, alternatives to the existing paradigms are intensively sought after. Recently, the proposal of two-dimensional (2D) ferrovalley materials has laid out a magneto-valleytronics composite paradigm based on spontaneous valley polarization induced by the integrated effects of magnetic order and spin-orbit coupling (SOC), which can radically reduce additional costs of the applied external fields[4]. Specifically, when valley couples with intrinsic ferromagnetic order, the valley-dependent Berry phase effects can generate emergent valley anomalous transport phenomena — e.g., valley Hall effect[5–11] and valley Nernst effect[12,13]) — making it possible to realize high-performance quantum devices and thus raising an intensive interest in materials systems which host magneto-valley traits.

From the perspective of potential applications in valleytronics, exploring various phase transitions in 2D magneto-valley materials has played a vital role in promoting our understanding, discovery, and characterization of emerging quantum states of matter. A general law of phase transitions is that they can drive not only rich quantum states but also intriguing physical properties. However, possibly due to the fewer number of 2D ferrovalley materials[4,14–18] and the difficulty in characterizing the magnetic ordering and topological phase transitions, high-quality magneto-valley material candidates with a wide range of phase transitions have been scarcely discussed. On the other hand, an interplay between different phase transitions accompanied by distinct valley anomalous transport manifestations in such materials has not been seriously considered yet, which poses a great challenge for the research on potential high-performance valleytronics devices.

A crucial but thought-provoking issue for magneto-valley coupling is finding a way to effectively generate spontaneous valley polarization utilizing the spin degree of freedom, and thereby lead to a revolution in magneto-valley-based information storage and operation principles. To tackle the current challenges of valleytronics, we break out the present ferrovalley paradigm to provide a general classification of magneto-valley coupling states through an effective model analysis, where the low energy electronic inter-valley nature enjoys a giant valley splitting with fully spin-polarized fermions. Consistent with the model, as a concrete example, $VSi_2N_4$ showcases emerging magneto-valley coupling phases — including valley-half-semiconductor, valley-half-semimetal as well as valley-unbalanced quantum anomalous Hall states (see Fig. 1) — transitions between which can be controlled by using external stimuli such as biaxial strain, electric field, and correlation effects. Remarkably, we find that topological phase transitions with magneto-valley coupling can exhibit valley-dependent Berry phase effects which manifest in prominent valley sign-reversible anomalous transport fingerprints.

## RESULTS AND DISCUSSION

### Valley-half-semiconductors and their topological phase transitions

According to the relationship between the valence band maximum and conduction band minimum of the same spin channel at the Fermi level, magneto-valley coupling (MVC) gives rise to a wide spectrum of quantum phase transitions ranging from gapped to gapless ones. To better capture the key physics underlying topological phase transitions, we construct a simple tight-binding (TB) model for describing the topological phase transitions of MVC states. Without loss of generality, we take a 2D triangular lattice with a magnetic space group $P\bar{6}m'2'$ as an example, and assume that the direction of the spins of magnetic atoms is parallel to the z-axis (i.e., out of the plane). To conveniently describe the atomic basis of $P\bar{6}m'2'$ symmetry, the minimal set of $|d_{x^2-y^2}\uparrow\rangle$ and $|d_{xy}\uparrow\rangle$ (or $|p_x\uparrow\rangle$ and $|p_y\uparrow\rangle$) orbitals is taken as the basis in our TB model. The Hamiltonian containing only the nearest-neighbour hopping reads:

$$H = \sum_{\langle ij\rangle\alpha\beta} t_{ij}^{\alpha\beta} c_{i\alpha}^\dagger c_{j\beta} + H.c., \quad (1)$$

where $c_{i\alpha}^\dagger$ ($c_{j\beta}$) is the electron creation (annihilation) operator for the orbital $\alpha$ ($\beta$) on site $i$ ($j$). By using the MagneticTB package[19] that was recently developed to incorporate the group representation theory, all independent nearest-neighbour hopping integrals $t_{ij}^{\alpha\beta}$ can be screened under a given symmetry constraint. Under the magnetic space group of $P\bar{6}m'2'$, the topological phase transitions of MVC states can be achieved with only three parameters [$r_1$, $r_2$, and $e_1$, see Eqs. (23) and (24)]. The Chern number for characterizing topologically nontrival phases can be analytically written as,

$$|\mathcal{C}| = \left|\frac{1}{2}\left(\text{Sign}\left(e_1 - \frac{3\sqrt{3}r_2}{2}\right) - \text{Sign}\left(e_1 + \frac{3\sqrt{3}r_2}{2}\right)\right)\right| \quad (2)$$

[1]Centre for Quantum Physics, Key Laboratory of Advanced Optoelectronic Quantum Architecture and Measurement (MOE), School of Physics, Beijing Institute of Technology, 100081 Beijing, China. [2]Beijing Key Lab of Nanophotonics & Ultrafine Optoelectronic Systems, School of Physics, Beijing Institute of Technology, 100081 Beijing, China. [3]College of Mathematics and Physics, Beijing University of Chemical Technology, 100029 Beijing, China. [4]Peter Grünberg Institut and Institute for Advanced Simulation, Forschungszentrum Jülich and JARA, 52425 Jülich, Germany. [5]Institute of Physics, Johannes Gutenberg University Mainz, 55099 Mainz, Germany. [6]These authors contributed equally: Xiaodong Zhou, Run-Wu Zhang, Zeying Zhang. ✉email: wxfeng@bit.edu.cn

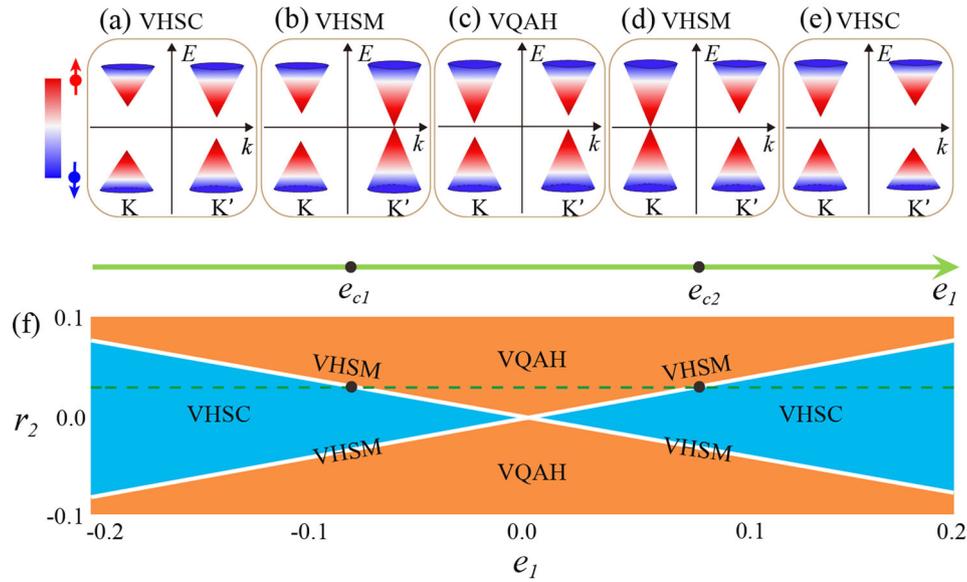

**Fig. 1 Schematic diagrams of valley-dependent topological phase transitions.** The balanced system is a valley-half-semiconductor (VHSC) state (**a**), which simultaneously exhibits valley polarized, full spin-polarized, and semiconducting properties. With increasing the parameter $e_1$, K′ and K valleys close successively at the critical points $e_{c1}$ and $e_{c2}$ respectively, corresponding to the valley-half-semimetal (VHSM) states (**b**, **d**), which simultaneously exhibits valley polarized, full spin-polarized, and semi-metallic properties. The fully spin-polarized valley-unbalanced quantum anomalous Hall states (VQAH) state (**c**) is expected between two VHSM states, which harbors valley index and quantum anomalous Hall effect simultaneously. Beyond $e_{c2}$, the gap at K valley reopens and the VHSC state restores (**e**). A phase diagram as a function of the model parameters $r_2$ and $e_1$ ($r_1 = 2$) is summarized in (**f**).

For our model, Fig. 1a schematically shows that K and K′ valleys bare semiconductor characteristics with full spin-polarization in the same spin channel. Such a valley-half-semiconductor (VHSC) state is highly promising for generating, transporting, and manipulating spin currents in spin-valleytronics. Topological phase transitions starting from the VHSC state can be realized by changing the model parameters $r_2$ and $e_1$ simultaneously. When the bandgap at K valley decreases, the other one at K′ valley is reduced as well. During the transition, a critical state, namely the valley-half-semimetal (VHSM) state, is inevitably encountered [Fig. 1b]: it is gapless at K′ valley but it is gapped at K valley. The gapless crossing point is two-fold degenerate with linear dispersion, similar to that of Weyl semimetals. In Fig. 1c, the gap reopening at K′ valley indicates a topological phase transition, and the valley-unbalanced quantum anomalous Hall state (VQAH) state is confirmed by a nonzero Chern number ($\mathcal{C} = 1$), which is contributed by both K and K′ valleys. The VQAH state predicted here differs from the previously considered quantized valley Hall state or valley-polarized quantum anomalous

Hall state. The quantized valley Hall state is found in spatial inversion broken non-magnetic materials, in which both K and K′ valleys host nonzero Chern numbers but with opposite signs, while the valley-polarized quantum anomalous Hall state is predicted in magnetic materials with nonzero Chern number originating from only one valley[5,20,21]. Further increasing the parameter $e_1$ will force K valley to first close the gap and then reopen it again, as shown in Fig. 1d, e. The summary of topological phase transitions is shown in Fig. 1f. Note that the SOC has been taken into account in our model, and therefore the emergence of MVC states having 100% spin polarization is remarkable and promising for spin-valleytronics[22–25].

An interplay between valley degree of freedom, magnetism, and topology provides an excellent platform for researching valley-related anomalous transport properties of MVC states, realizing a rich set of exotic quantum phenomena, such as valley anomalous Hall effect (VAHE), valley anomalous Nernst effect (VANE), as well as valley magneto-optical Kerr effect (VMOKE) and valley magneto-optical Faraday effect (VMOFE). Specifically, pursuing a single material that simultaneously exhibits topological phase transitions and valley sign-reversible Berry phase effects has been rarely done to date, although such a material would be highly valuable for multi-functional miniaturized devices. In addition to considering artificially constructed MVC heterostructures, an alternative is to seek intrinsic MVC materials that can harbor topological phase transitions driven by external means, such as biaxial strain, electric field, and correlation effects.

### High-quality candidate materials

In this work not only do we provide a classification of MVC states, but also propose a series of MVC materials [Fig. 2a, b], including 2D ferromagnetic $VSi_2N_4$ as well as other eleven $MA_2Z_4$ ($M$ = V, Nb; $A$ = Si, Ge; $Z$ = N, P, As) candidates[26,27]. All these candidate materials form on a hexagonal lattice with the same magnetic space group that has been employed above, which can be regarded as the magnetic counterparts of the valley Hall semiconductors $MoSi_2N_4$ family[28,29]. While aforementioned MVC states and their topological phase transitions exist in all materials, below we focus on one representative — VSi$_2$N$_4$ — and discuss the rest of the materials in the Supplemental Material (see Supplementary Figs. 1 and 2).

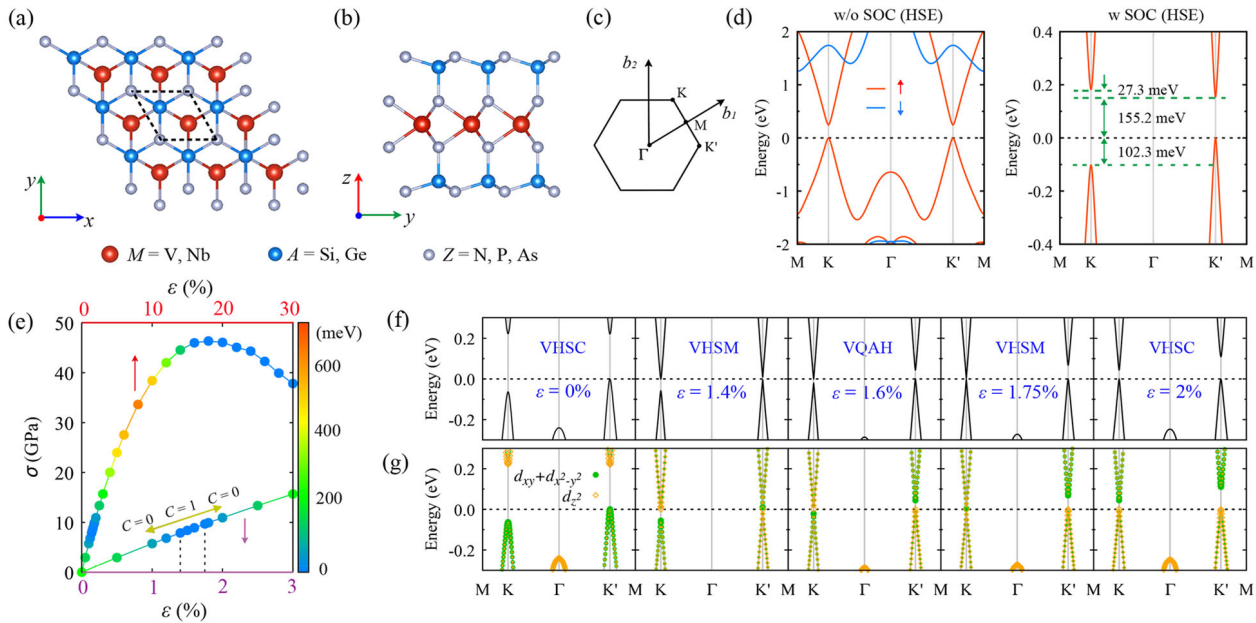

**Fig. 2 Valley-dependent topological phase transitions in VSi$_2$N$_4$. a, b** The top and side views of $MA_2Z_4$ materials family. The dashed lines in (**a**) indicate the 2D primitive cell. **c** The Brillouin zone of $MA_2Z_4$. **d** The band structures of VSi$_2$N$_4$ were calculated by the hybrid functional without SOC (left panel) and with SOC (right panel). (**e**) The stress-strain curve of VSi$_2$N$_4$. The top x-axis plots the whole strain range of 0–30%, while the bottom x-axis plots the small strain range of 0–3%. The colors indicate the magnitudes of the bandgap, and the corresponding Chern number ($C$) is marked. The evolution of band structure (**f**) and orbital compositions at band edge (**g**) under the strain of 0–2%. Various fully spin-polarized valley topological states emerge, including VHSC (0%), VHSM (1.4%), VQAH (1.6%), VHSM (1.75%), and VHSC (2%).

The electronic band structure of VSi$_2$N$_4$, which is a fully spin-polarized semiconductor with a small direct bandgap (0.24 eV) at K/K' valley in the spin-up channel and with a large indirect bandgap (3.19 eV) in the spin-down channel, is shown in Fig. 2d. When SOC is taken into account, VSi$_2$N$_4$ evolves into a VHSC state with a giant valley splitting of 102.3 (27.3) meV at the valence (conduction) band edge, which is originated from the simultaneous breaking of time-reversal and spatial inversion symmetries. Its physical origin is similar to the resonant photovoltaic effect in doped magnetic semiconductors[30]. More importantly, VSi$_2$N$_4$ possesses an "ultra-clean" linear band dispersion around K and K' valleys in the energy range of [−0.7, +1.2] eV, which suppresses the influence of irrelevant trivial bands for valley performance. The linear band dispersion ensures a high Fermi velocity of about $0.4 - 0.5 \times 10^6$ m/s along with different momentum directions, which is of the same order as that of graphene[31]. This promotes VSi$_2$N$_4$ into one of the most appealing candidates when comparing it with currently considered ferrovalley materials. Note that the VHSC state of VSi$_2$N$_4$ has been previously discovered in a recent work[32].

While VSi$_2$N$_4$ is a long-sought VHSC material, it can also host various fully spin-polarized MVC topological states under external means such as biaxial strain, electric field, and correlation effects. Taking biaxial strain as an example (the cases of electric field and correlation effects are shown in Supplementary Figs. 5 and 6), we predict strained VSi$_2$N$_4$ to exhibit various MVC states, covering VHSC, VHSM, and VQAH states. As shown in Fig. 2f, by increasing strain within a reasonable range (0–2%), the VHSC state (0%) undergoes topological phase transitions that can be identified by the closing and reopening of band gaps at K' and K valleys, resulting in the VHSM (1.4%), VQAH (1.6%), VHSM (1.75%), and VHSC (2%) states. The gap evolution together with corresponding Chern numbers is summarized in Fig. 2e. Furthermore, the orbital-projected band structures of MVC states [Fig. 2g] are another indicator for topological phase transitions. In the balanced state of VSi$_2$N$_4$, the valence band maximum is dominated by the $d_{xy}$ and $d_{x^2-y^2}$ orbitals of V atoms, and the conduction band minimum mainly comes from the $d_{z^2}$ orbital. At the strain of 1.4%, the orbital compositions reverse at K' valley, driving the system into the VQAH state; further increasing strain to 1.75%, the orbital inversion occurs also at K valley, restoring the system to the VHSC state. This valley-related topological phase transition can be also driven by the correlation effect, as realized in FeCl$_2$[17] and VSi$_2$P$_4$[33,34].

### Sign reversal of valley-dependent Berry phase effects
An in-depth investigation of MVC topological phase transitions provides a platform for exploring valley-related anomalous transport phenomena. In this context, it is remarkable that VSi$_2$N$_4$, while hosting topological phase transitions, exhibits also sign change of valley-dependent Berry phase effects. Regarding the VHSC state (0%), for which the calculated **k**-resolved Berry curvature **Ω(k)** is shown in Fig. 3a, one can clearly identify the hot spots in the Berry curvature around two valleys with opposite signs and different magnitudes. By introducing a tiny biaxial strain, the VHSC state experiences a topological phase transition into the VQAH state, bridged by the VHSM state. Within the strain of 1.4–1.6%, the sign of **Ω(k)** at K' valley flips [Fig. 3b]. Further increasing strain from 1.6–2%, K valley also experiences a topological phase transition, akin to the case of K' valley, resulting in the sign change of **Ω(k)** at K valley [Fig. 3c]. Such dynamics of **Ω(k)** is bound to influence valley-related anomalous transport phenomena such as VAHE, VANE, VMOKE, and VMOFE.

Our predictions concerning the valley-related anomalous transport phenomena are presented in Fig. 3. Since VAHE is calculated by the integration of **Ω(k)** in a small region centered at each valley, the sign changes of VAHE are in full accordance with **Ω(k)** during the topological phase transitions. This is also the case for VANE as well as VMOKE and VMOFE. The former is calculated by integrating the Berry curvature together with a weighting factor around each valley (see Eq. (4)). The latter phenomenon can be actually regarded as the *ac* counterpart of VAHE. Physically, these anomalous transport phenomena are intimately related to each other. Therefore, the change in their signs strongly depends on the nature of topological phase transitions, exhibiting exotic sign reversal of valley-dependent Berry phase effects.

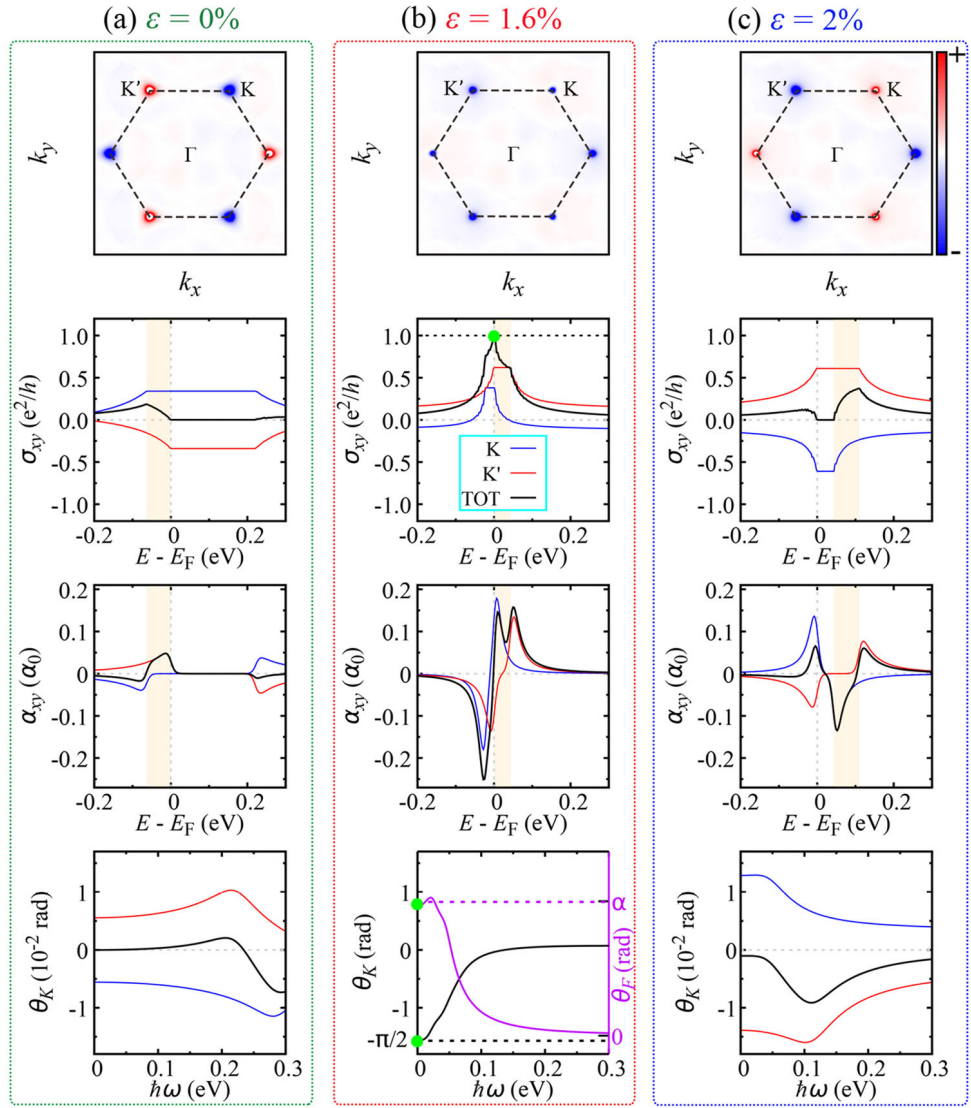

**Fig. 3 Sign-reversible valley-dependent anomalous transport properties in VSi$_2$N$_4$.** The Berry curvature $\Omega(\mathbf{k})$ (first row), VAHE (second row), VANE (third row), and VMOKE (fourth row) under three different strains of 0% (**a**), 1.6% (**b**), and 2% (**c**), respectively. The results of VMOFE (not shown here) are rather similar to that of VMOKE. The unit of VANE is set to be $a_0 = k_B e/h$ (e is the elementary charge, $k_B$ and $h$ are Boltzmann and Planck constants, respectively). Yellow shaded areas indicate the regions of large valley splitting. The Berry curvature is plotted at the energy where the anomalous Hall conductivity reaches its maximum. The legends of K, K′, and TOT represent contributions from K valley, K′ valley, and the entire Brillouin zone, respectively.

Interestingly, due to the different magnitudes of $\Omega(\mathbf{k})$ at two valleys, a net fully spin-polarized valley current is produced by the anomalous Hall, anomalous Nernst, and magneto-optical effects. Notably, the VQAH phase emerges during the phase transition, accompanied by a quantized anomalous Hall conductivity $\sigma_{xy} = e^2/h$ [second row of Fig. 3b]. Besides, one can utilize the non-contact magneto-optical technique[35–37] to detect this topological phase[38–41]. For a VQAH state, the central physical idea is that Maxwell's equations have to be modified by adding an axion term $(\Theta\alpha/4\pi^2)\mathbf{E}\cdot\mathbf{B}$ (here $\Theta$ is magnetoelectric polarizability and $\alpha = e^2/\hbar c$ is the fine structure constant) into the usual Lagrangian[42]. In this way, the magneto-optical Kerr ($\theta_K$) and Faraday ($\theta_F$) rotation angles turn out to be quantized in the low-frequency limit, that is, $\theta_K \simeq -\pi/2$ and $\theta_F \simeq \mathcal{C}\alpha$ ($\mathcal{C}$ is the Chern number)[43,44]. In the bottom of Fig. 3b, one can clearly observe the quantization behavior of $\theta_K$ and $\theta_F$ in the low-frequency limit.

Generally, the tunable sign of $\Omega(\mathbf{k})$ has been witnessed by reversing the magnetization[4] and ferroelectric polarization[45]. However, these two means seem to be sub-optimal and often suffer from various drawbacks, making it difficult to utilize this effect. As an alternative avenue, the manipulation by a small biaxial strain is more suitable for practical purposes. It is worth noting that topological phase transitions through biaxial strain may not only change the sign of anomalous transport characteristics but also modify their magnitude, and more importantly, the mediated quantization of transport characteristics is indispensable for experimental observation. We further calculate the effects of temperature and disorder on the valley-dependent Berry phase effects, and the results (see Supplementary Fig. 10) show that the sign changes of VAHE and VANE, being the key feature of topological phase transitions, are robust against disorder and temperature.

In this work, we introduce a general framework to realize topological magneto-valley phase transitions in 2D VHSC, and we propose a series of feasible candidate materials harboring valley-dependent Berry phase effects, which are triggered by external means such as biaxial strain, electric field, and correlation effects. Taking $VSi_2N_4$ as a representative, we demonstrate that such intrinsic VHSC states display long-sought fully spin-polarized valley index. The proposal of sign reversal of valley-dependent Berry phase effects and high-quality materials realization greatly expand the ferrovalley family and provide an exciting playground for spintronics and valleytronics applications.

## METHODS
### First-principles calculations

The first-principles calculations were carried out employing the projected augmented wave method[46], as implemented in the Vienna ab initio simulation package (VASP)[47]. The exchange-correlation effect was treated by the Perdew-Burke-Ernzerhof parameterized generalized-gradient approximation (PBE-GGA)[48]. The energy cut-off of 500 eV and the k-mesh of $25 \times 25 \times 1$ were used in the static calculations. The force and energy convergence criteria were set to be $10^{-3}$ eV/Å and $10^{-7}$ eV, respectively. The on-site Coulomb correlation of V and Nb atoms is considered within the GGA + U scheme[49]. Different effective Hubbard energy U were tested in Supplementary Fig. 5 and the U = 3 eV was used in the main text, which has been also used in ref. [32]. The more accurate Heyd-Scuseria-Ernzerhof hybrid functional method (HSE06)[50] was used to check the electronic structure. A vacuum space of 13 Å was used to avoid the interactions between the neighboring slabs. The phonon spectrum was performed based on the density functional perturbation theory (DFPT)[51]. The mostly localized Wannier functions including the d-orbitals of V atom, the s- and p-orbitals of Si atom, and the p-orbitals of N atom were constructed on a k-mesh of $8 \times 8 \times 1$, using the WANNIER90 package[52].

### Anomalous Hall and anomalous Nernst effects

The intrinsic anomalous Hall conductivity (AHC) and anomalous Nernst conductivity (ANC) were calculated on a dense k-mesh of $501 \times 501 \times 1$, using the Berry phase theory[53,54],

$$\sigma_{xy} = -\frac{e^2}{\hbar} \sum_n \int \frac{d^2k}{(2\pi)^2} \Omega_{xy}^n(\mathbf{k}) w_n(\mathbf{k}), \tag{3}$$

$$\alpha_{xy} = \frac{e k_B}{\hbar} \sum_n \int \frac{d^2k}{(2\pi)^2} \Omega_{xy}^n(\mathbf{k}) W_n(\mathbf{k}), \tag{4}$$

where $\Omega_{xy}^n(\mathbf{k})$ is the band- and momentum-resolved Berry curvature

$$\Omega_{xy}^n(\mathbf{k}) = -\sum_{n' \neq n} \frac{2\mathrm{Im}[\langle \psi_{n\mathbf{k}} | \hat{v}_x | \psi_{n'\mathbf{k}} \rangle \langle \psi_{n'\mathbf{k}} | \hat{v}_y | \psi_{n\mathbf{k}} \rangle]}{(\omega_{n'\mathbf{k}} - \omega_{n\mathbf{k}})^2}. \tag{5}$$

Here, $\{x, y\}$ denote the Cartesian coordinates, $\hat{v}_{x,y}$ are the velocity operators, and $\psi_{n\mathbf{k}}$ ($\hbar\omega_{n\mathbf{k}} = E_{n\mathbf{k}}$) is the eigenvector (eigenvalue) at band index n and momentum $\mathbf{k}$. The two unitless weighting factors $w_n(\mathbf{k})$ and $W_n(\mathbf{k})$ in Eqs. (3) and (4) are written as

$$w_n(\mathbf{k}) = f_n(\mathbf{k}), \tag{6}$$

$$W_n(\mathbf{k}) = \frac{1}{k_B T} [(E_{n\mathbf{k}} - \mu) f_n(\mathbf{k}) + k_B T \ln(1 + e^{-(E_{n\mathbf{k}} - \mu)/k_B T})], \tag{7}$$

where $f_n(\mathbf{k}) = 1/[\exp((E_{n\mathbf{k}} - \mu)/k_B T) + 1]$ is the Fermi-Dirac distribution function, T is temperature, $\mu$ is chemical potential, and $k_B$ is the Boltzmann constant.

Physically, the anomalous Hall effect can be influenced by disorder[55–57]. To model the variation of intrinsic anomalous Hall conductivity against disorder, a constant $\Gamma$ approximation of the Kubo formula was used[55],

$$\begin{aligned}
\sigma_{xy} = &-\frac{e^2 \hbar}{2\pi} \int \frac{d^2k}{(2\pi)^2} \sum_{mn, m \neq n} \mathrm{Im}\{v_{mn}^x(\mathbf{k}) v_{nm}^y(\mathbf{k})\} \\
&\times \Big\{ \frac{(E_{m\mathbf{k}} - E_{n\mathbf{k}})\Gamma}{[(E_F - E_{n\mathbf{k}})^2 + \Gamma^2][(E_F - E_{m\mathbf{k}})^2 + \Gamma^2]} \\
&- \frac{2\Gamma}{(E_{m\mathbf{k}} - E_{n\mathbf{k}})[(E_F - E_{m\mathbf{k}})^2 + \Gamma^2]} \\
&+ \frac{2}{(E_{m\mathbf{k}} - E_{n\mathbf{k}})^2} \mathrm{Im}(\ln \frac{E_F - E_{m\mathbf{k}} + \mathrm{i}\Gamma}{E_F - E_{n\mathbf{k}} + \mathrm{i}\Gamma})\Big\},
\end{aligned} \tag{8}$$

where $v^j$, $E_F$, and $\Gamma$ are the velocity operator, Fermi energy, and band broadening parameter, respectively. The anomalous Nernst conductivity under the effect of the disorder can be then calculated through the generalized Mott relation,

$$\alpha_{xy} = -\frac{1}{e} \int dE \frac{\partial f}{\partial \mu} \sigma_{xy} \frac{E - \mu}{T}, \tag{9}$$

where f, E, $\mu$, T, and $\sigma_{xy}$ are the Fermi-Dirac distribution function, energy, chemical potential, temperature, and anomalous Hall conductivity, respectively. When $\Gamma \to 0$, the anomalous Hall and Nernst conductivities calculated from Eqs. (8) and (9) have to be identical to the results of Eqs. (3) and (4), respectively.

### Magneto-optical Kerr and Faraday effects

Extending the AHC to the ac case, the optical Hall conductivity[52],

$$\begin{aligned}
\sigma_{xy}(\omega) = &\frac{\mathrm{i}e^2 \hbar}{N_k V_c} \sum_{\mathbf{k}} \sum_{n,m} \frac{f_{m\mathbf{k}} - f_{n\mathbf{k}}}{E_{m\mathbf{k}} - E_{n\mathbf{k}}} \\
&\frac{\langle \psi_{n\mathbf{k}} | \hat{v}_x | \psi_{m\mathbf{k}} \rangle \langle \psi_{m\mathbf{k}} | \hat{v}_y | \psi_{n\mathbf{k}} \rangle}{E_{m\mathbf{k}} - E_{n\mathbf{k}} - (\hbar\omega + \mathrm{i}\eta)},
\end{aligned} \tag{10}$$

was evaluated on a k-mesh of $1001 \times 1001 \times 1$. Here, $V_c$ is the cell volume, $N_k$ is the number of k-points, $\hbar\omega$ is the photon energy, and $\eta$ is the smearing parameter.

The MO Kerr and Faraday effects in topologically trivial two-dimensional (2D) magnetic materials are given by[58–61],

$$\theta_K = \mathrm{Re}\left[\mathrm{i} \frac{2\omega d}{c} \frac{\sigma_{xy}}{\sigma_{xx}^s}\right], \tag{11}$$

$$\theta_F = \mathrm{Re}\left[\frac{\omega d}{2c}(n_+ - n_-)\right], \tag{12}$$

here $\theta_K$ and $\theta_F$ are the MO Kerr and Faraday rotation angles, respectively. c is the speed of light in vacuum, $\hbar\omega$ is the photon energy, d is the effective thickness of the 2D magnetic material, $\sigma_{xx}^s$ is the diagonal element of the optical conductivity for a non-magnetic substrate (the $SiO_2$ is used, and the substrate effect was discussed in our previous works[62,63]), $\sigma_{xy}$ is the off-diagonal element of the optical conductivity of the 2D magnetic materials, and $n_\pm = [1 + \frac{4\pi\mathrm{i}}{\omega}(\sigma_{xx} \pm \mathrm{i}\sigma_{xy})]^{1/2}$ are the eigenvalues of the dielectric tensor of the 2D magnetic materials.

The MO Kerr and Faraday rotation angles in the Chern insulator are written by[64],

$$\theta_K = \frac{1}{2}\left(\arg\{E_+^r\} - \arg\{E_-^r\}\right), \tag{13}$$

$$\theta_F = \frac{1}{2}\left(\arg\{E_+^t\} - \arg\{E_-^t\}\right), \tag{14}$$

in which $E_\pm^{r,t} = E_x^{r,t} \pm \mathrm{i} E_y^{r,t}$ are the left (+) and right (−) circularly polarized components of the reflected (r) and transmitted (t) outgoing electric fields. When the thickness of the Chern insulators is much thinner than the incident light wavelength, the reflected and transmitted fields can be obtained[64],

$$E_x^r = [1 - (1 + 4\pi\sigma_{xx})^2 - (4\pi\sigma_{xy})^2]A, \tag{15}$$

$$E_y^r = 8\pi\sigma_{xy}A, \tag{16}$$

$$E_x^t = 4(1 + 2\pi\sigma_{xx})A, \tag{17}$$

$$E_y^t = 8\pi\sigma_{xy}A, \tag{18}$$

with $A = 1/[(2 + 4\pi\sigma_{xx})^2 + (4\pi\sigma_{xy})^2]$. In the low-frequency limit ($\omega \to 0$), the optical conductivity in Chern insulators gives $\sigma_{xx}^R = 0$, $\sigma_{xx}^I = 0$, $\sigma_{xy}^R = \mathcal{C}e^2/h$, and $\sigma_{xy}^I = 0$ ($\mathcal{C}$ is the Chern number; the superscripts R and

$I$ are the real and imaginary parts, respectively). Then, the Eqs. (13)–(18) can be simplified to[43,44,64],

$$\theta_K = -\tan^{-1}[c/(2\pi\sigma_{xy}^R)] = -\tan^{-1}(1/\mathcal{C}a) \simeq -\pi/2, \quad (19)$$

$$\theta_F = \tan^{-1}(2\pi\sigma_{xy}^R/c) = \tan(\mathcal{C}a) \simeq \mathcal{C}a. \quad (20)$$

Thus, in the low-frequency limit, the magneto-optical Kerr and Faraday rotation angles in Chern insulators are quantized to $\theta_K = -\pi/2$ and $\theta_F = \mathcal{C}a$, respectively.

### Curie temperature

Regarding the ferromagnetic VSi$_2$N$_4$, a key physical quantity is Curie temperature ($T_c$), which can be estimated by carrying out the Monte Carlo simulations based on the Ising Hamiltonian model[65],

$$H = -\sum_{ij} J_{ij} \mathbf{S}_i \cdot \mathbf{S}_j. \quad (21)$$

Here, $J_{ij}$ is the nearest-neighboring exchange interaction, $\mathbf{S}$ is the spin magnetic moment on the V or Nb atom. The calculated $T_c$ for strain-free ($\varepsilon = 0\%$) VSi$_2$N$_4$ is about 100 K [see Supplementary Fig. 3b], which is more than two times larger than CrI$_3$ of 45 K[36] and Cr$_2$Ge$_2$Te$_6$ of less than 30 K[37], indicating the potential applications of 2D spintronics. The $T_c$ (75 K) of strained structure ($\varepsilon = 1.6\%$) is slightly smaller than that of the balanced system but is much larger than that of the famous antiferromagnetic topological insulator MnBi$_2$Te$_4$ (24 K)[66], suggesting the great application prospects to realize the high-temperature VQAH state.

### Tight-binding model

The magnetic space group of monolayer VSi$_2$N$_4$ with out-of-plane magnetization is $P\bar{6}m'2'$. Considering the magnetic V atoms, it is a 2D triangular lattice with out-of-plane ferromagnetic order. We now introduce a two-band tight-binding model using a minimal basis set of $|d_{x^2-y^2}\uparrow\rangle$ and $|d_{xy}\uparrow\rangle$ (or $|p_x\uparrow\rangle$ and $|p_y\uparrow\rangle$) orbitals, where ↑ means spin up. Then, the generators of the magnetic space group $P\bar{6}m'2'$ are represented by $S_{3z} = e^{2\pi i \sigma_y/3 - \pi i \sigma_0/6}$ and $M_x \mathcal{T} = -i\sigma_z$, respectively. By including only the nearest-neighbour hopping, the symmetry-adapted tight-binding Hamiltonian can be written by,

$$H = \sum_{\langle ij \rangle \alpha\beta} t_{ij}^{\alpha\beta} c_{i\alpha}^\dagger c_{j\beta} + H.c., \quad (22)$$

where $c_{i\alpha}^\dagger$ ($c_{j\beta}$) is the electron creation (annihilation) operator for the orbital $\alpha$ ($\beta$) on-site $i$ ($j$). Note that this Hamiltonian is in general including spin-orbit coupling and magnetic interaction. By using the MagneticTB package[19] that is recently developed on the top of group representation theory, all independent nearest-neighbour hopping integrals $t_{ij}^{\alpha\beta}$ can be screened under a given symmetry constrain. Under magnetic space group of $P\bar{6}m'2'$, we found there are only three parameters are necessary to realize the magneto-valley coupling states and their topological phase transitions. The Hamiltonian is recast to

$$\begin{aligned} H(\mathbf{k}) = & (e_1 + r_2(\sin\tilde{k} - \sin k_x - \sin k_y))\sigma_y \\ & + r_1(\cos\tilde{k} - \cos k_y)\sigma_x \\ & - r_1 \frac{(\cos\tilde{k} - 2\cos k_x + \cos k_y)}{\sqrt{3}} \sigma_z, \end{aligned} \quad (23)$$

where $\tilde{k} = k_x + k_y$, $r_1 = t_1 + t_2$, $r_2 = t_1 - t_2$, and $e_1$ are real parameters. Specifically, $t_1$ and $t_2$ are the nearest-neighbor hopping integrals and $e_1$ is the on-site hopping integral from different orbitals. When $r_2 \ll r_1$, such a model with different bandgaps at K and K′ valleys are actually a modified Haldane's model[67]. We can further expand this Hamiltonian around K and K′ valleys:

$$H^\pm = \frac{1}{2}\sqrt{3}r_1(k_x + 2k_y)\sigma_x + \left(\pm e_1 - \frac{3\sqrt{3}r_2}{2}\right)\sigma_y - \frac{3k_x r_1}{2}\sigma_z, \quad (24)$$

where $H(K) = H^+$ and $H(K') = -H^-$. The Chern number for characterizing topological nontrival phases can be analytically written as,

$$|\mathcal{C}| = \left| \frac{1}{2}\left( \text{Sign}\left(e_1 - \frac{3\sqrt{3}r_2}{2}\right) - \text{Sign}\left(e_1 + \frac{3\sqrt{3}r_2}{2}\right) \right) \right|. \quad (25)$$

## DATA AVAILABILITY
The data that support the findings of this study are available from the corresponding author on reasonable request.

## CODE AVAILABILITY
The first principle and tight-binding codes that support the findings of this study are available from the corresponding author on reasonable request.

Received: 11 July 2021; Accepted: 11 September 2021;
Published online: 01 October 2021

## ACKNOWLEDGEMENTS

This work is supported by the National Key R&D Program of China (Grant No. 2020YFA0308800), the National Natural Science Foundation of China (Grant Nos. 11734003, 11874085, 12047512, and 12004028), and the Project Funded by China Postdoctoral Science Foundation (Grant Nos. 2020M680011 and 2021T140057). Y.M. acknowledges the Deutsche Forschungsgemeinschaft (DFG, German Research Foundation) − TRR 288 - 422213477 (project B06). Y.M., W.F., and Y.Y. acknowledge the funding under the Joint Sino-German Research Projects (Chinese Grant No. 12061131002 & German Grant No. 1731/10-1) and the Sino-German Mobility Programme (Grant No. M-0142).

## AUTHOR CONTRIBUTIONS

W.F. and Y.Y. conceived the research. X.Z., and R.-W.Z. performed the first-principles calculations. Z.Z. carried out the model analysis. X.Z., R.-W.Z., and Z.Z. contributed equally to this work. All authors contributed to the discussion of the data. X.Z., R.-W.Z., W.F., and Y.M. wrote the paper with discussion from and Y.Y.

## COMPETING INTERESTS

The authors declare no competing interests.

## ADDITIONAL INFORMATION

**Supplementary information** The online version contains supplementary material available at https://doi.org/10.1038/s41524-021-00632-3.

**Correspondence** and requests for materials should be addressed to Wanxiang Feng.

**Reprints and permission information** is available at http://www.nature.com/reprints

**Publisher's note** Springer Nature remains neutral with regard to jurisdictional claims in published maps and institutional affiliations.